\documentclass[aps,superscriptaddress,eqsecnum,nofootinbib,showpacs,preprintnumbers]{revtex4}
\usepackage{graphicx,epsfig}
\usepackage{amsmath}
\usepackage {amssymb}

\newcommand{\be}{\begin{eqnarray}}
\newcommand{\ee}{\end{eqnarray}}
\newcommand{\bea}{\begin{eqnarray}}

\newcommand{\eea}{\end{eqnarray}}

\newcommand{\no}{\nonumber}

\begin{document}

\title{Entanglement Entropy as a Probe of the Proximity Effect in Holographic Superconductors}
\author{Xiao-Mei Kuang}
\email{kuangxiaomei@sjtu.edu.cn} \affiliation{INPAC, Department of
Physics and Shanghai Key Lab for Particle Physics and Cosmology,
Shanghai Jiao Tong University, Shanghai 200240, China.}
\author{Eleftherios Papantonopoulos}
\email{lpapa@central.ntua.gr} \affiliation{Department of Physics,
National Technical University of Athens, GR-15780 Athens, Greece.}
\affiliation{INPAC, Department of Physics and Shanghai Key Lab for
Particle Physics and Cosmology, Shanghai Jiao Tong University,
Shanghai 200240, China.}
\author{Bin Wang}
\email{wang_b@sjtu.edu.cn} \affiliation{INPAC, Department of
Physics and Shanghai Key Lab for Particle Physics and Cosmology,
Shanghai Jiao Tong University, Shanghai 200240, China.}

\date{\today}

\begin{abstract}

We study the entanglement entropy as a probe of the proximity
effect of a superconducting system by using the gauge/gravity
duality in a fully back-reacted gravity system. While the
entanglement entropy in the superconducting phase is  less than
the entanglement entropy in the normal phase, we find that near
the contact interface of the superconducting to normal phase   the
entanglement entropy has a different behavior due  to the leakage
of Cooper pairs to the normal phase.   We verify this behavior by
calculating the conductivity near the boundary interface.

\end{abstract}

\date{\today}



\maketitle

\section{Introduction}
\label{sec:1}

The gauge/gravity duality has been proven to be a very powerful
tool in studying strongly coupled phenomena using dual
gravitational systems where the coupling is weak
\cite{Maldacena:1997re}. This duality which is well founded in
string theory has many interesting  applications and among them
one which from the first sight it seems unexpected, in condensed
matter physics (for a review, see \cite{Hartnoll:2009sz}). A
condensed matter system that is well studied using the
gauge/gravity duality is the holographic superconductor.

The simplest holographic superconductor model which is extensively
studied is described by an Einstein-Maxwell-scalar field theory
with a negative cosmological constant
\cite{Hartnoll:2008vx,Hartnoll:2008kx}. Its gravity sector is
described by an Abelian-Higgs model with a stationary black hole
metric  which  below a certain critical temperature the black hole
acquires scalar hair. Its dual boundary field theory  is described
by a theory which is similar to the standard Landau-Ginzburg
theory in which the scalar field corresponds to an operator, the
order parameter, which condenses below the critical temperature,
signalizing the onset of superconductivity.

This phenomenological approach has the virtue of simplicity, but
does not capture all the underlying features of the gauge/gravity
duality including quantum effects. This is hard to implement, as
it remains a challenge to embed this model in a quantum system
(string/M-theory). In spite of that, this holographic principle
 has been applied to many other
condensed  matter systems like   the  conventional and
unconventional superfluids and superconductors
\cite{Gubser:2008wv}, Fermi liquid behavior
\cite{Cubrovic:2009ye}, non-linear hydrodynamics
\cite{Bhattacharyya:2008jc}, quantum phase transitions
\cite{Iqbal:2010eh} and transport \cite{Herzog:2007ij}. In all
these  studies it is crucial to understand the low temperature
limit. In the gravity sector this requires to go beyond the probe
limit and to consider fully back-reacted gravitational systems. In
the boundary theory the $T\rightarrow 0$ limit leads to strongly
coupled systems in the quantum physics regime where new superfluid
phenomena arise like for example the generation of inhomogeneous
FFLO phases \cite{Alsup:2012ap,Alsup:2012kr}. Recently it was
proposed \cite{Albash:2012pd} that this low temperature regime can
be probed by the entanglement entropy
\cite{Ryu:2006bv,Ryu:2006ef}.

The entanglement entropy can be considered as a measure of how a
given quantum system is strongly correlated (entangled). It was
introduced as a tool to describe  different phases and their
corresponding phase transitions of a quantum system as the
temperature goes to zero. The entanglement entropy is directly
related to the degrees of freedom of a system, keeping track of
them. It can also play the role of  an order parameter of a
 phase transition at very low temperatures. The most important
 property of the entanglement entropy is that it is non-vanishing at zero
temperature. For these reasons it was employed as a probe of
quantum properties of the ground state for a given quantum system.

Recently the entanglement entropy was used to study various
properties of holographic superconductors at low temperatures. In
a model coming from $\mathcal{N}=8$ gauged supergravity
\cite{Albash:2012pd} the entanglement entropy across the
superconducting phase transition was studied. It was found that
the entanglement entropy is lower in the superconducting phase
than in the normal phase. This behavior was attributed to some
kind of reorganization of the degrees of freedom of the system:
because electrons are bounded in the superconducting phase to form
Cooper pairs less degrees of freedom remain comparable to the
normal phase.

The behavior of entanglement entropy across the holographic p-wave
superconductor phase transition in an
 Einstein-Yang-Mills theory with a negative cosmological constant was
 studied in
 \cite{Cai:2012nm,Cai:2012sk}. In \cite{Cai:2013oma}  a holographic p-wave superconductor/insulator model
 was considered and it was found that   as the back reaction
increases, the transition is changed from second order to  first
order. Also in  \cite{Arias:2012py}   the gravitational
backreaction of the non-Abelian gauge field on the gravity dual to
a 2+1 $p$-wave superconductor was studied. It was  found that the
$p$-wave superconductor has lower entanglement entropy. The
   behavior of holographic entanglement
entropy   for imbalanced holographic superconductor was considered
in \cite{Dutta:2013osl}. It was found that entanglement entropy
for this imbalanced system decreases with the increase of
imbalance in chemical potentials.

In this work we will use the entanglement entropy as a probe of
the proximity
 effect   in a holographic superconductor.
In condensed matter physics the proximity effect describes the
dynamics of a system near the superconductor-normal metal
interface where  the superconducting electrons (Cooper pairs) may
penetrate from the superconducting to normal phase. The leakage of
the Cooper pairs weakens the superconductivity near the interface
with a normal metal. This phenomenon can appear even in the
absence of a magnetic field \cite{Buzdin}. One of our motivations
in this work is to construct a computationally tractable gravity
model and show that it can reproduce basic properties of proximity
effect in superconductivity.

The phenomenological analysis of the proximity effect was given by
the generalized Ginzburg-Landau theory \cite{Buzdin}. In this
theory a complex scalar field $\Psi$ is considered as the
superconducting order parameter which contributes to the free
energy in the functional \be
F_G=a(T)|\Psi|^2+\frac{b(T)}{2}|\Psi|^4+\gamma(T)|\overrightarrow{\nabla}\Psi|^2~,
\ee where the coefficient $a$ vanishes at the transition
temperature $T_c$. At $T<T_c$, the coefficient $a$ is negative and
the minimum of $F_G$ occurs for a uniform superconducting state
with $|\Psi|^2=-a/b$. The coefficient $a$ is given by
$a=\alpha(T-T_c)$, where $T_c$ is the critical temperature of the
transition into the uniform superconducting state. The appearance
of the proximity effect can simply be interpreted as the effect of
the gradient term in the Ginzburg-Landau functional.

Let us consider the decay of the order parameter in the normal
phase, i.e., at $T>T_c$ assuming that our system is in contact
with another superconductor with a higher critical temperature,
and the $x$ axis is chosen perpendicular to the interface between
the superconductor and the normal phases. The induced
superconductivity is weak and, we use the linearized
Ginzburg-Landau equation for the order parameter,
$a\Psi-\gamma\frac{\partial^2\Psi}{\partial x^2}=0$, with $\gamma
\neq 0$. The decaying solution is $\Psi=\Psi_0\exp[-x/\xi(T)]$,
where $\xi(T)=\sqrt{\gamma/a}$ is the correlation length. When
$\gamma>0$, the order parameter decays exponentially as
$\Psi(\gamma>0)=\Psi_0\exp[-x/\xi(T)]$. But when $\gamma<0$, close
to the interface, the order parameter $\Psi(\gamma<0)\sim
\Psi_0\cos(x/\sqrt{|\gamma|/a})$, which is bigger than
$\Psi(\gamma>0)$ in the normal phase close to the interface of the
superconductor. This shows that due to the gradient term, the
superconducting properties can be induced in the normal phase.
This phenomenon is called the proximity effect. Simultaneously the
leakage of the Cooper pairs weakens the superconductivity in the
superconductor phase near the interface with a normal metal, which
results in a decrease of the superconducting transition
temperature in a thin superconducting layer in contact with a
normal metal.  Using the generalized Ginzburg-Landau functional
\cite{Buzdin}, we see that the gradient term plays an important
role in inducing the proximity effect.

Our aim here is to build  a holographic superconductor with a dual
boundary field described by a theory similar to the generalized
Ginzburg-Landau theory. Such a theory, in the probe limit, was
discussed in \cite{kuang:2013a} in which a  $U(1)$ gauge field was
introduced along with a complex scalar field coupled to a charged
AdS black hole and a higher-derivative coupling between the $U(1)$
gauge field and the scalar with coupling constant $\eta$. This
coupling is provided by a potential term which in a covariant form
reads $V(\Psi)=m^2|\Psi|^2+\eta
|F^{\mu\nu}D_\nu\Psi|^2,\label{holpot} $ where $F=dA$ is the
strength of a U(1) gauge field $A_\mu$, $\Psi$ is a charged
complex scalar field of charge $q$ and mass $m$ and $D_\mu
=\nabla_\mu-iqA_\mu$. However, this covariance is broken on the
boundary, because the coefficient $\eta$ plays the role of
$\gamma$ of the boundary Ginzburg-Landau theory and both of them
indicate the strength of the gradient derivatives of the scalar
field. What we had found in \cite{kuang:2013a} is that, in the
probe limit, large positive values of the coupling $\eta$ make
easier the transition to a superconducting phase, while when
$\eta$ becomes negative, it is more difficult for the condensation
to be formed and this happens because the energy gap in the probe
limit  is larger for $\eta <0$ than the energy gap in the
conventional case ($\eta =0$).

One can expect that such high-derivatives terms can arise
 in  string theory. Indeed, there
are models based on
 exact solutions of $D=11$ and type IIB
supergravity in which after  consistent  Kaluza-Klein truncations
to four spacetime dimensions, fully back-reacted solutions
describing holographic superconductors in three spacetime
dimensions have been found
\cite{Gauntlett:2009zw,Gauntlett:2009dn}. These models,  contain a
large number of scalar, gauge fields and high derivatives of them,
which need to be constrained in order to make the models tractable
\cite{Gauntlett:2009bh,Bobev}.

    It
is of great interest to generalize our previous study to the fully
back-reacted theory. More interestingly, we would like to
construct a tractable gravity model to reproduce the proximity
effect in the holographic superconductor. Our $\eta$, which is
analogous to $\gamma$ in the general Ginzburg-Landau functional
\cite{Buzdin}, gives us the hope to build a holographic
description of the proximity effect in the gravity model. Using
the fact that the
  entanglement entropy
 plays the role of an order parameter, we will calculate it near
 the interface of
 superconducting/normal phase of a holographic superconductor in low temperatures
 and we will show that it  gives us  important information
 on the behavior of system.  To verify this behavior we will also
calculate the conductivity near the boundary interface.

The work is organized as follows. In Section~\ref{sec:2} we
present the gravity sector of the model. In Section~\ref{sec:3} we
present the fully back-reacted solution of the holographic system.
In Section~\ref{sec:4} we discuss the entanglement entropy. In
Section~\ref{sec:5} we calculate the conductivity and finally in
Section~\ref{sec:6} are our conclusions.

\section{The gravitational sector}
\label{sec:2}

We will consider a scalar field coupled to a $U(1)$ gauge field
with an action discussed in \cite{kuang:2013a}
\begin{equation}\label{action}
 S=\int d^4x \sqrt{-g}\Big[\frac{R+6/L^2}{16\pi
 G}-\frac{1}{4}F_{\mu\nu} F^{\mu\nu}
 -|D_{\mu}\Psi|^2-V(\Psi)\Big]~,
\end{equation}
where the potential term is given by the expression above.
 The field equations  are given by:
\begin{itemize}
\item the Einstein equations
\begin{eqnarray}\label{0eqEinstein}
R_{\mu\nu}-\frac{1}{2}Rg_{\mu\nu}-\frac{3}{L^2}g_{\mu\nu}=8\pi
GT_{\mu\nu}~,
\end{eqnarray}
\item the Maxwell equations \begin{eqnarray} \label{0eqMaxwell}
 &\nabla_{\mu}&F^{\mu\nu}+\frac{\eta}{\sqrt{-g}}\partial_\mu\Big[\sqrt{-g}(D_\kappa \Psi)(D_\lambda
 \Psi)^*\Big(g^{\kappa\nu}F^{\mu\lambda}-g^{\kappa\mu}F^{\nu\lambda}+g^{\nu\lambda}F^{\mu\kappa}-
 g^{\mu\lambda}F^{\nu\kappa}\Big)\Big]\nonumber\\
 &=&iq\Big[\Psi^* (D^\nu \Psi)-\Psi(D^\nu
 \Psi)^*\Big]+iq\eta g_{\mu\rho}F^{\rho\nu}\Big[F^{\mu\kappa}\Psi^*
 (D_\kappa\Psi)-F^{\mu\lambda}\Psi
 (D_\lambda\Psi)^*\Big],
\end{eqnarray}
\item and the scalar field equation
\begin{eqnarray} \label{0eqPsi}
&-&\frac{1}{\sqrt{-g}}\partial_\mu
\bigl[\sqrt{-g}g^{\mu\nu}\bigl(\partial_\nu\Psi-iqA_\nu
\Psi\bigr)\bigr]+iqg^{\mu\nu}A_\nu\Big(\partial_\mu\Psi-iqA_\mu\Psi\Big)+
m^2\Psi\nonumber
\\&=&\frac{\eta}{\sqrt{-g}}\partial_\mu \Big[\sqrt{-g}
g_{\kappa\lambda}F^{\kappa\nu}F^{\lambda\mu}\Big(\partial_\nu\Psi-iqA_\nu
\Psi\Big)\Big]-iq\eta
g_{\kappa\lambda}F^{\kappa\nu}F^{\lambda\mu}A_\mu
\Big(\partial_\nu\Psi-iqA_\nu \Psi\Big)~.
\end{eqnarray}
\end{itemize}
In this work, we will set $L=1$, $8\pi G=1$, $q=1$.

We note that the presence of the coupling constant $\eta$ adds new
terms in the field equations which makes the system of the
differential equations highly non-trivial. We have to find
numerical solutions of the fully back-reacted system.

\section{The solutions of the holographic system}
\label{sec:3}

We will generalize the probe limit discussion in
\cite{kuang:2013a} to a full back-reacted formalism by taking the
ansatz of metric and matter fields as
\begin{equation}\label{metric1}
d s^2 = - \frac{1}{z^2} f(z) e^{- \chi(z) } dt^2 + \frac{1}{z^2}
\Big( dx^2 + dy^2 \Big) + \frac{1}{z^2} \frac{d z^2}{f(z)} \
 , ~~~A_{\mu} = A_t(z)dt  \ ,  ~~~\Psi = \psi(z) \ .
\end{equation}
The temperature can be expressed as
\begin{equation}\label{temperature}
T=\frac{-e^{\chi/2}\partial_zf}{4\pi}|_{z=z_h}~.
\end{equation}
Then, the independent Einstein-Maxwell equations reduced from
(\ref{0eqEinstein}) and (\ref{0eqMaxwell}) become
\begin{eqnarray}
\label{EE1}
0&=&\chi'-\frac{z \Big(1+2 e^{\chi } z^4 \eta  \text{At}'^2\Big) \Big(e^{\chi }q^2 \text{At}^2 \psi ^2+f^2 \psi '^2\Big)}{f^2}~,\\
\label{EE2}
0&=&f'+\frac{3(1-f)}{z}-\frac{m^2\psi^2}{2z}-\frac{e^{\chi}z^3A_t^{'2}}{4}-\frac{e^{\chi}zq^2A_t^{2}
(2+5e^{\chi}z^4\eta A_t^{'2})\psi^2}{4f}-\frac{zf(2+3e^{\chi}\eta
z^4A_t^{'2})}{4}\psi^{'2}~,\\
\label{1eqMax} 0&=&A_t''+\Big[\frac{z^2f^2\chi'+4e^{\chi}\eta
z^3\psi q^2A_t^2(z\psi f'-2f(\psi+z\psi'+\frac{3}{4}z\psi\chi'))}
{2 a_0 f}\no\\&+&\frac{2\eta z^3f\psi'(z
f'\psi'+2f(\psi'(1+\frac{1}{4}z\chi')+z\psi''))}{a_0}\Big]A_t'
-\Big[\frac{2\psi^2q^2(1+e^{\chi}\eta z^4A_t'^2)}{a_0}\Big]A_t~,
\end{eqnarray}
while the scalar field equation (\ref{0eqPsi}) takes the form
\begin{eqnarray}
\label{1eqpsi}
0&=&\psi''+\Big[\frac{f'}{f}+\frac{(4+z\chi')(1+e^{\chi}z^4\eta
A_t^{'2})+4z^5e^{\chi}\eta A_t^{'}A_t^{''}}{2z(e^{\chi} z^4\eta
A_t^{'2}-1)}\Big]\psi'+\Big[\frac{e^{\chi}q^2A_t^2}{f^2}+\frac{m^2}{z^2f(e^{\chi}
z^4\eta A_t^{'2}-1)}\Big]\psi~,
\end{eqnarray}
with $a_0=z^2f(1+2\eta z^2f\psi'^2)-2e^{\chi}\eta
z^4\psi^2q^2A_t^2$. Here (\ref{EE1}) and (\ref{EE2}) are the
combination of $tt$ and $zz$ component of Einstein's equation and
the xx component can be obtained from differentiating the two
Einstein equations above.

Near the boundary $z\rightarrow0$, we need $\chi(z=0)$=0 to
recover the pure AdS boundary. The matter fields should behave as
\begin{eqnarray}\label{asypsiAt1}
\psi&=&\psi^{(1)}z^{\Delta_{-}}+\psi^{(2)}z^{\Delta_{+}}+\cdots,\\
A_t&=&\mu-\rho z+\cdots,
\end{eqnarray}
where according to the AdS/CFT dictionary, $\psi^{(i)}=\langle
O_{i}\rangle/\sqrt{2},~i=1,2$ and $O_{i}$ with the conformal
dimensions $\Delta_\pm=\frac{3}{2}\pm\frac{1}{2}\sqrt{9+4 m^2}$
are the corresponding dual operators of $\psi^{(i)}$ in the field
theory side.  $\mu$ and $\rho$ are the corresponding chemical
potential and charge density in the dual boundary field theory,
respectively. In this paper, we focus on the case $m^2=-2$ and set
$\psi^{(2)}=0$ to consider $\psi^{(1)}$ as the vacuum expectation
value of the operator $\langle O_{1}\rangle$.

At the horizon $z=z_h$, the regular condition implies $A_t(z_h)=0$
and $f(z_h)=0$. Then we can expand all the fields near the horizon
and use the scaling symmetries
\begin{eqnarray}\label{scalesymmetry}
&&t\rightarrow a t,\quad z \rightarrow az ,\nonumber\\
&&z \rightarrow az, \quad (t,x,y) \rightarrow (t,x,y)/a ,  \quad
A_t \rightarrow a A_t,\quad f\rightarrow f,\quad \psi\rightarrow
\psi,
\end{eqnarray}
to set $z_h=1$.

At high temperature, the scalar field will vanish and there is no
condensation. The solution is the Reissner-Nordstr\"{o}m AdS black
hole with
\begin{eqnarray}
\chi = \psi = 0 \,,\qquad A_t =  \mu \Big(1-\frac{z}{z_h}\Big)\,,
\qquad f = 1 - \frac{z^3}{z_h^2} \Big(1 + \frac{\mu^2z_h^2}{4}
\Big) + \frac{z^4}{z_h^4} \Big(\frac{\mu^2z_h^2}{4} \Big)~.\
\end{eqnarray}
The temperature is give by
\begin{eqnarray}
T=\frac{1}{4\pi z_h}\Big(3-\frac{\mu^2z_h^2}{4}\Big)~.
\end{eqnarray}

\begin{figure}[h]
\center{
\includegraphics[scale=0.6]{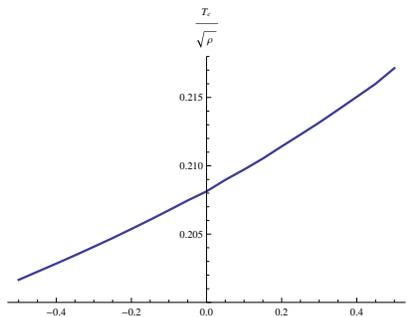}\hspace{0.3cm}
\caption{The dependence of critical temperature on the coupling
constant $\eta$.} \label{fig-eta-Tc}}
\end{figure}

\begin{figure}[h]
\center{
\includegraphics[scale=0.55]{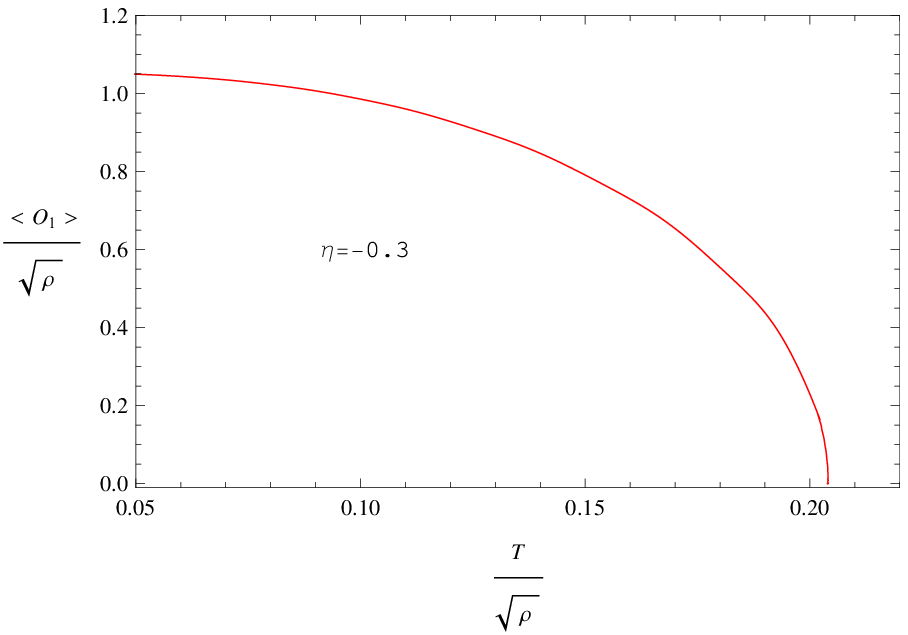}\hspace{0.3cm}
\includegraphics[scale=0.55]{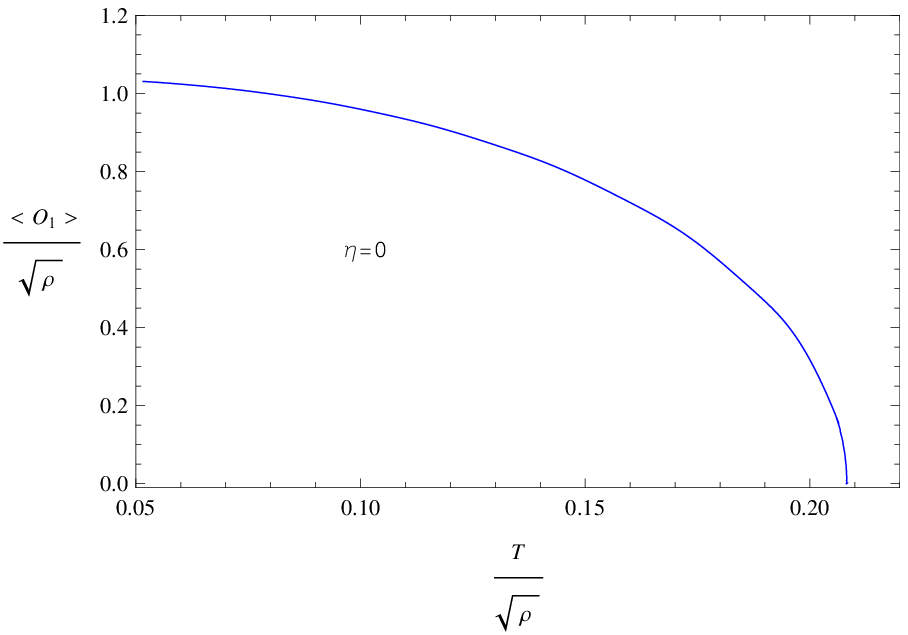}\hspace{0.3cm}
\includegraphics[scale=0.55]{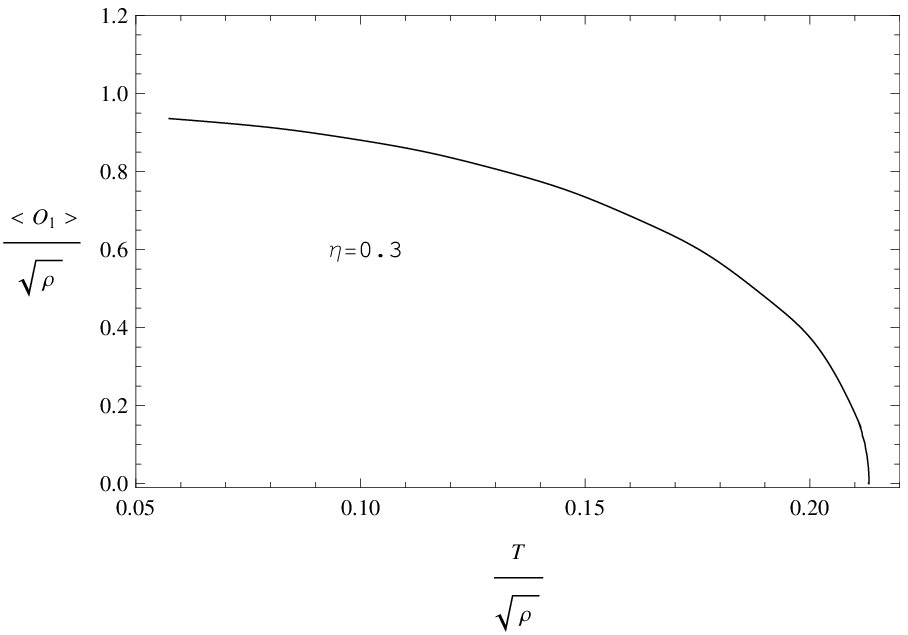}\hspace{0.3cm}
\caption{Plots of the operator $O_1$ versus temperature after
condensation with different values of the coupling constant
$\eta$.} \label{fig-condensation}}
\end{figure}

When the temperature decreases to be lower than a critical value,
a new type of charged black hole with non-vanishing charged scalar
profile is numerically available. This corresponds to a hairy
phase with  $O_1$ non-vanishing. The explicit dependence of the
critical temperature for $O_1$ on the  coupling is presented in
Fig.~\ref{fig-eta-Tc}. We see that the critical temperature
increases as the  coupling $\eta$ becomes larger, which is
consistent with the results in the probe limit in our previous
work \cite{kuang:2013a}. Moreover, we show the vacuum expectation
values for $O_1$ in Fig.~\ref{fig-condensation}. It is observed
that as the coupling increases, the condensation gap is lower
which agrees well with the property of the critical temperature we
discussed above. Thus, from the gravitational side, we have found
that the greater strength of the interaction will make the
condensation easier to form in the back-reacted background. In the
dual boundary field theory, this means that with stronger coupling
between the $U (1)$ gauge field and the scalar field, the gauge
symmetry can be broken more easily.

%
%
Having the solutions of the normal phase and the hairy phase below
the critical temperature, we will study the holographic
entanglement entropy (HEE) from high to lower temperatures in the
next Section.

\section{The entanglement entropy of the holographic system}
\label{sec:4}

Let us now discuss how we can incorporate the notion of
entanglement entropy in the AdS/CFT correspondence. Imagine that
we have a system $A$ in  the boundary  CFT  which has a gravity
dual. Since the information included in a subsystem $B$ is
evaluated by the entanglement entropy $S_A$, we can ask which part
of AdS space is responsible for the calculation of $S_A$ in the
dual gravity side. In  \cite{Ryu:2006bv} a formula was proposed
 \begin{equation}\label{RTF}
S_A=\frac{\mbox{Area}(\gamma_A)}{4G^{(d+2)}_N}\ ,
\end{equation}
where $\gamma_A$ is the $d$-dimensional minimal surface whose
boundary is given by the $(d-1)$-dimensional manifold $\partial
\gamma_A=\partial A$.
The constant $G^{(d+2)}_N$ is the Newton constant of the general
gravity in AdS$_{d+2}$.


Before studying the holographic entanglement entropy (HEE) in this
holographic model, we will geometrize the HEE of (\ref{RTF}) in
terms of AdS/CFT duality. Following \cite{Albash:2012pd}, we will
consider the subsystem $A$ with a straight strip geometry
described by $-\frac{l}{2}\leq x \leq \frac{l}{2}\,, 0\leq y \leq
L~,$ where $l$ is defined as the size of region $A$ and $L$ is a
regulator which can be set to be infinity.
%
The induced metric of the hypersurface $\gamma_A$ whose boundary
is the same as the stripe and has a profile like (\ref{metric1})
reads as
\begin{equation}
ds_{induced}^2=\frac{1}{z^2}\Big[(\frac{1}{f}+x'(z))dz^2+dy^2\Big]~.
\end{equation}
Thus, the HEE connecting with the area of the surface can be
expressed as
\begin{equation}\label{Area}
4G_4S=Area(\gamma_A)=L\int_{-l/2}^{l/2}\frac{dx}{z^2}\sqrt{1+\frac{z'(x)^2}{f}}~.
\end{equation}
The above expression can be treated as the Lagrangian with $x$
direction thought of as time. The corresponding Hamiltonian is
conserved because the Lagrangian does not explicitly depend on
$x$. So we can get a constant of motion as
\begin{equation}\label{zstar}
\frac{1}{z_*^2}=\frac{1}{z^2\sqrt{1+\frac{z'(x)^2}{f}}}~,
\end{equation}
with $z_{*}$ satisfying the condition $dz/dx|_{z=z_{*}}=0$. From
(\ref{zstar}), we can write the width $l$ in terms of $z_{*}$
\begin{equation}\label{length}
\frac{l}{2}=\int_{\epsilon}^{z_{*}}dz\frac{z^2}{\sqrt{(z_{*}^4-z^4)f}}~.
\end{equation}
Substituting (\ref{zstar}) into (\ref{Area}), we can obtain the
entanglement entropy
\begin{equation}\label{HEE}
4G_4S=2L\int_{\epsilon}^{z_{*}}dz\frac{z_{*}^2}{z^2}\frac{1}{\sqrt{(z_{*}^4-z^4)f}}=2L(s+\frac{1}{\epsilon})~.
\end{equation}
Here the term $1/\epsilon$ is divergent. While the term $s$ is a
finite term, so it is physically important.

Now, we can calculate the entanglement entropy $s$ based on the
solution discussed in the last section. We will see the behavior
of HEE from normal phase to hairy phase and investigate the effect
of the $\eta$ coupling. Note that according to the scaling
symmetries in (\ref{scalesymmetry}), the dimensionless quantities
are $T/\sqrt{\rho}$, $\sqrt{\rho} l$ and $s/\sqrt{\rho}$.
\begin{figure}[h]
\center{
\includegraphics[scale=0.5]{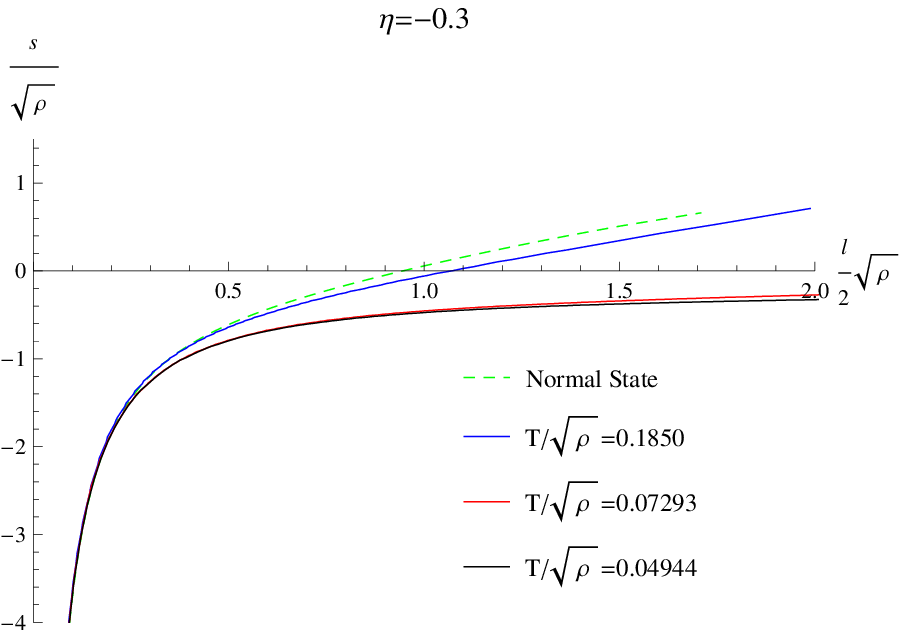}\hspace{0.5cm}
\includegraphics[scale=0.5]{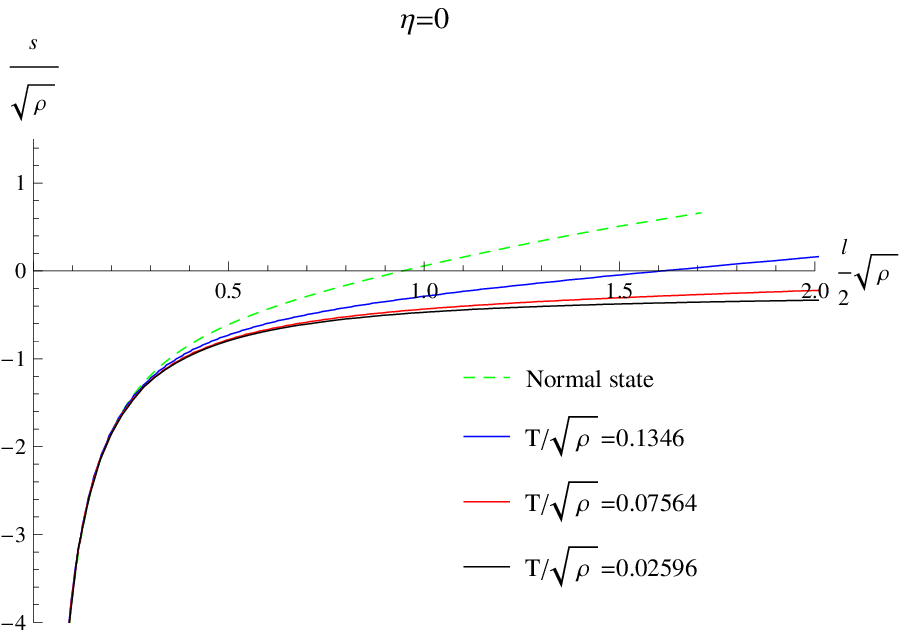}\hspace{0.5cm}
\includegraphics[scale=0.5]{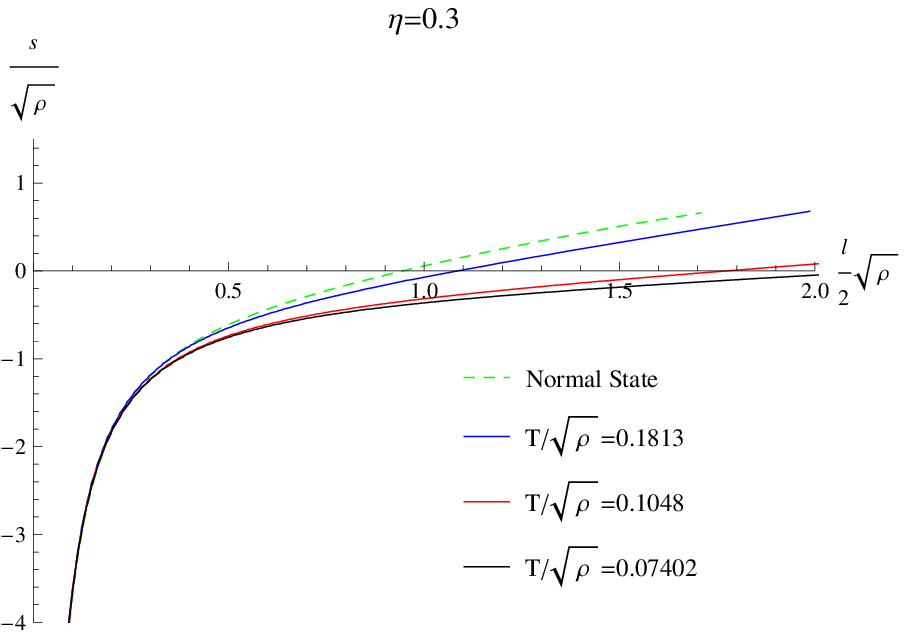}\hspace{0.5cm}
\caption{The entanglement entropy as a function of stripe width
for various temperatures and different values of the coupling $
\eta$.} \label{fig-l-s_ChangT}}
\end{figure}

Fixing the temperature, we see in Fig.~\ref{fig-l-s_ChangT} how
the HEE changes as the width of stripe $l$ changes . The green
line is for the normal phase with the RN-AdS black hole
background. We see that below the critical temperature when the
scalar field starts to condensate, the entanglement entropy
becomes smaller and it drops when the temperature becomes lower.
This is consistent with the expectation that in the
superconducting phase the degrees of freedom decrease due to the
formation of Cooper pairs \cite{Albash:2012pd}. This property
holds for different values of the coupling $\eta$.

\begin{figure}[h]
\center{
\includegraphics[scale=0.8]{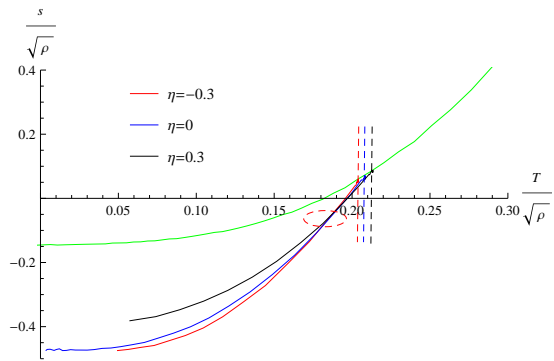}\hspace{0.5cm}
\caption{The entanglement entropy as a function of temperature for
a fixed $\sqrt{\rho}\frac{l}{2}=1$. The green curve depicts
 the normal phase while the other curves depict the superconducting phases
 with different values of the coupling $\eta$. The region in the
ellipse is enlarged in Fig.~\ref{fig-T-sL}.} \label{fig-T-s}}
\end{figure}

To illustrate the influence of the $\eta$ coupling, we present the
HEE in change of temperature for a fixed $l$ in
Fig.~\ref{fig-T-s}. The light green line is the HEE for the normal
state with RN-AdS black hole background. As the temperature
decreases, the slope of HEE presents a discontinuous change at a
critical temperatures $T_c$ denoted by vertical dashed lines in
the figure for different strength of the coupling $\eta$. The
discontinous change of the HEE marks the phase transition point
from the normal state to the superconducting state. We again
observe that the superconducting phase always have smaller HEE
after the phase transition.

At low temperature, the HEE  of a superconductor with a smaller
$\eta$  is smaller. Physically this happens  because  at low
temperature the condensation becomes stronger with higher
condensation gap for smaller coupling $\eta$ so  the number of
Cooper pairs is increased, which results in  less degrees of
freedom available.

\begin{figure}[h]
\center{
\includegraphics[scale=0.8]{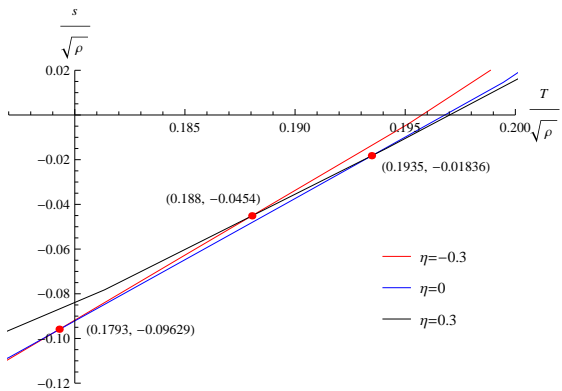}\hspace{0.5cm}
\caption{The region in the ellipse in Fig.~\ref{fig-T-s} is shown
here enlarged. As exhibited in the plot, the curves of HEE for the
coupling $\eta=0$ intersects with that for $\eta=-0.3$ and
$\eta=0.3$ at $T/\sqrt{\rho}=0.1793$ and $T/\sqrt{\rho}=0.1935$,
respectively. While the curves  with the couplings $\eta=-0.3$ and
$\eta=0.3$ intersect at $T/\sqrt{\rho}=0.188$. This sharp change
of HEE near the interface is due to the proximity effect.}
\label{fig-T-sL}}
\end{figure}

However this property does not hold when the temperature increases
near the critical value, which is close to the interface with the
normal phase,  marked in the ellipse in Fig.~\ref{fig-T-s}, which
is enlarged in Fig.~\ref{fig-T-sL}. When the temperature is
increased above $T/\sqrt{\rho} = 0.1793$, we see that the HEE for
$\eta= -0.3$ becomes higher than the case with $\eta = 0$. When
the temperature is above $T/\sqrt{\rho} = 0.188$, the HEE for
$\eta = -0.3$ surpasses the value for $\eta= 0.3$. The sharp
change of the HEE which is related to the coupling $\eta$ in the
vicinity of the contact interface, is not trivial. This behavior
can be attributed to the proximity effect.

For the smaller coupling holographic superconductor, it is more
likely that the Cooper pairs of the supercontacting state can
penetrate  the normal state, which effectively results in the
increase of the HEE in the superconductor phase. This is a similar
behavior to the   generalized Ginzburg-Landau theory on the
boundary, where the phenomenological coefficient $\gamma$ is
related to the correlation length $\xi=\sqrt{\gamma/|a|}$ as we
discussed in the introduction. A more negative $\gamma$ for a
fixed $a$ leads more Cooper pairs to penetrate  and this results
in a larger leftover order parameter in the normal phase
\cite{Buzdin}. On the other hand, for larger
 positive values of the coupling $\eta$ coupling, Cooper pairs remain in
the superconductor phase, thus leads the HEE to relatively smaller
values. Again  this agrees to the generalized Ginzburg-Landau
theory description that more positive $\gamma$ leads to smaller
penetration of the Cooper pairs so that the leftover order
parameter in the normal phase is weaker.

\section{conductivity}
\label{sec:5} In this section we will discuss the conductivity in
an attempt to understand better the behavior of the system near
the critical temperature. In the superconductor, the conductivity
possesses certain distinguishing properties which are largely
dependent on the microscopic details. To support the results we
found studying the entanglement  entropy, we will calculate the
real part of conductivity for two characteristic temperatures at
which we observed two different behaviors of the entanglement
entropy. The first one is a  low temperature which characterizes
the superconducting phase where the entanglement entropy decreases
with the decrease of the coupling constant $\eta$, at which more
Cooper pairs are formed for more negative coupling $\eta$. For a
high  temperature we choose a temperature at which the
superconducting state had just been formed  and effectively it is
close to the transition temperature  of the superconducting to
normal state. At the high temperature we found that the
entanglement entropy is higher for negative $\eta$.

We consider the perturbation of metric and $U(1)$ field as $\delta
g_{\mu\nu}=g_{tx}(z)e^{-i\omega t}$ and $\delta
A_{\mu}=A_x(z)e^{-i\omega t}$. Then, the first order perturbation
equation of $g_{tx}$ can be deduced  as
\begin{equation}
g_{tx}'-\frac{2 g_{tx}}{z}-A_x A_t'\Big[1- z^2 \eta\Big(f
\psi'^2-\frac{q^2 e^{\chi} A_t^2 \psi^2}{f}\Big)\Big]=0~.
\end{equation}
Having this equation  we can deduce the linearised perturbative
Maxwell equation which decouples from $g_{tx}$
\begin{equation}
(1+2z^2\eta
f\psi^{'2})A_x^{''}+\Big(\frac{f'}{f}-\frac{\chi'}{2}+a_1\eta\Big)A_x'+\Big[
\Big(\frac{\omega^2}{f^2}-\frac{z^2A_t^{'2}}{f}\Big)e^{\chi}-\frac{2q^2\psi^2}{z^2f}+a_2\eta+a_3\eta^2\Big]A_x=0
\end{equation}
with
\begin{eqnarray}
a_1&=&z\psi'\Big[z\psi'(4f'-f\chi')+4f(\psi'+z\psi'')\Big]~,\nonumber\\
a_2&=&\frac{e^{2\chi}q^2z^2A_t^2\psi^2}{f^2}\Big(z^2A_t^{'2}-\frac{2\omega^2}{f}\Big)\nonumber\\&&-
\frac{e^{\chi}}{f}\Big[2q^2z^2A_t\psi
A_t^{''}+zA_t^{'2}(2q^2\psi^2+z^2\psi'f)+q^2zA_t\psi A_t^{'}
(4\psi+4z\psi'+z\psi\chi')\Big]~,\nonumber\\
a_3&=&2e^{\chi}z^6A_t^{'2}f\Big[\psi'^4-2e^{\chi}\Big(\frac{qA_t\psi\psi'}{f}\Big)^2
+e^{2\chi}\Big(\frac{qA_t\psi}{f}\Big)^4\Big]~.\nonumber\\
\nonumber
\end{eqnarray}

Although that above  equation  seems very complicated, it can be
solved by imposing the ingoing boundary condition near the horizon
\begin{equation}\label{Axz1}
A_x(z\rightarrow z_h)\propto f^{\frac{-i\omega}{4\pi T}}
\end{equation}
where the temperature $T$ is defined in (\ref{temperature}). Near
the asymptotic AdS boundary, the behavior of perturbation $A_x$ is
\begin{equation}\label{Axz0}
A_x(z\rightarrow 0)=A_x^{(0)}+z A_x^{(1)}~.
\end{equation}
Then, by invoking the AdS/CFT duality, the conductivity of
holographic superconductivity can be expressed as~\cite{Hartnoll:2008vx}
\begin{equation}\label{conductivity}
\sigma=-\frac{iA_x^{(1)}}{\omega A_x^{(0)}}~.
\end{equation}
After solving the equation with the boundary condition
(\ref{Axz1}), we can numerically extract the asymptotic value of
$A_x$ to calculate the conductivity of our system.

\begin{figure}[h]
\center{
\includegraphics[scale=0.8]{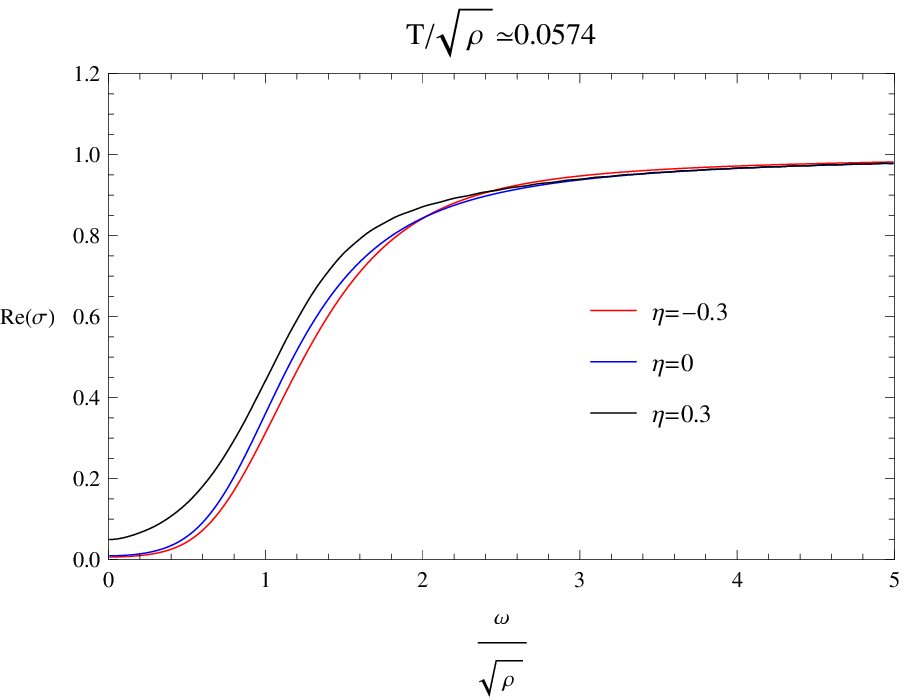}\hspace{0.5cm}
\includegraphics[scale=0.8]{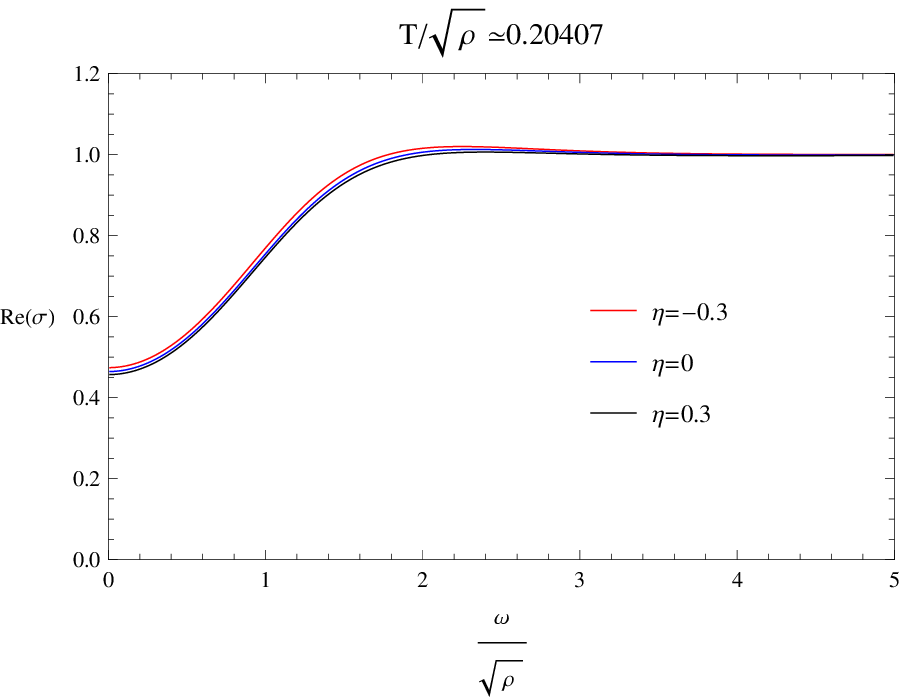}\hspace{0.5cm}
\caption{The real part of the conductivity with
$T=0.0574\sqrt{\rho}$ and $T=0.20407\sqrt{\rho}$.}
\label{fig-sigma1}}
\end{figure}

We concentrate on the real part of the conductivity,  since it is
the dissipative part of the conductivity and measures the presence
of charged states as a function of energy.  At the low
temperature, for example when $T/\sqrt{\rho}=0.0574$, which is
lower than the smallest crossing temperature in
Fig.~\ref{fig-T-sL}, we see in the left panel of
Fig.~\ref{fig-sigma1} that at low frequencies the holographic
superconducting system with smaller coupling $\eta$ has lower real
part of conductivity and a larger frequency gap.  The gap in the
 conductivity depends on the condensation,
$\omega_g\sim  <O>$,  which indicates a gap in the spectrum of
charged excitations. For the smaller coupling, the higher
condensation leads to the  larger superconducting gap and to a
smaller real part of the conductivity.  The drop in the real part
of the conductivity corresponds to a drop in the density of
excitations at energies below the chemical potential. Thus for
smaller $\eta$, more `electrons' are bounded in Cooper pairs which
explains the smaller HEE at low temperature.

\begin{figure}[h]
\center{
\includegraphics[scale=0.8]{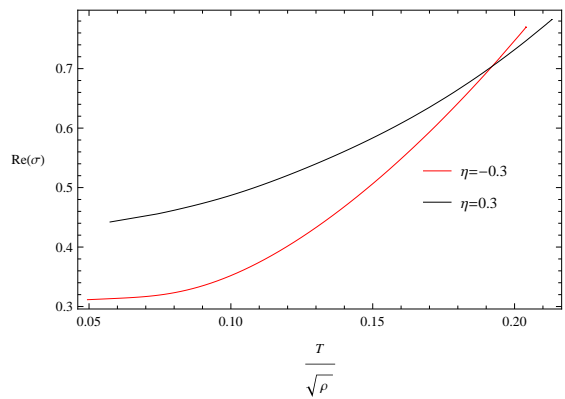}\hspace{0.5cm}
\caption{Dependence of the real part of the conductivity on the
temperature.} \label{fig-sigma2}}
\end{figure}

At high temperature near the critical point, where the dependence
of HEE on the coupling $\eta$ is reversed compared with the low
temperature case, the behavior of the real part of the
conductivity is plotted in the right plot of Fig.~\ref{fig-sigma1}
with $T/\sqrt{\rho}=0.20407$ for example, which is higher than the
largest crossing temperature in Fig.~\ref{fig-T-sL}. In the low
frequency, we see that the real part of conductivity is bigger
when the coupling is smaller. Fixing the frequency, we show the
dependence of the real part of the conductivity on the temperature
in Fig.~\ref{fig-sigma2}.  The sharp different dependence of the
real part of the conductivity on the coupling $\eta$ at low and
high temperatures are clearly shown.  In the contact interface,
 we see that for
the smaller $\eta$, the real part of the conductivity is higher.
The bigger real part of the conductivity in the contact interface
  supports the
proximity effect argument that less Cooper pairs are left in the
superconducting phase with smaller $\eta$ coupling.

\section{Conclusions}
\label{sec:6}

We have studied the behavior of the entanglement entropy of a
superconducting system.
Motivated by the generalized Ginzburg-Landau theory we have
introduced a higher-derivative coupling between the $U(1)$ gauge
field and the scalar field with coupling constant $\eta$. In the
boundary theory this coupling corresponds to one of the
phenomenological constants of the Landau-Ginzburg theory. We have
solved numerically the fully back-reacted gravitational system and
found that as the coupling $\eta$ increases the critical
temperature  increases and the energy gap  decreases. This result
suggests that the system with larger coupling is easier to enter
the superconducting phase.

Knowing the behavior of the system below the critical temperature,
we have calculated the entanglement entropy using a stripe
geometry. We have found that
the entanglement entropy is less than the entanglement entropy of
the normal phase. This agrees with the known results that the
 entanglement entropy plays the role of the order parameter of the
 superconducting system counting the degrees of freedom. In the
 superconducting phase less degrees of freedom are available
 because of the formation of Cooper pairs, therefore the entanglement
 entropy is less than the normal phase. 
 We also found that
 a larger coupling gives a larger entanglement entropy.
 This means
 that for a fixed temperature for a larger coupling, less Cooper
 pairs are available  in the superconducting system.

 We have found that near the boundary superconducting/normal
 interface the entanglement
 entropy has reversed behavior. More negative coupling gives higher entanglement
 entropy. This can be explained because of the proximity effect.
 For the more negative coupling, more Cooper pairs have leaked to the normal phase,
so that less electrons are bounded in the superconductor phase
which leads to the higher value of the entanglement
 entropy. This behavior of the entanglement
 entropy near the boundary contact interface is very
 interesting. It further
 supports the previous finding that the entanglement
 entropy plays the role of the order parameter, measuring the
 degrees of freedom of a superconducting system.

To support further this behavior of the entanglement
 entropy we have calculated the real part of conductivity
 for two characteristic temperatures, one low temperature and the
 other near the critical temperature which effectively close to the interface of the normal phase. For the low frequency at the low temperature,  smaller
 coupling has lower conductivity. But when the temperature is near the critical
 temperature,  the behavior of the conductivity is reversed due to
 the proximity effect.


In conclusion we have  built a holographic superconductor and
presented  a holographic description of the proximity effect in
 superconductivity. It would be interesting to extend this
study in the presence of an external magnetic field. For an
inhomogeneous magnetic field  the entanglement entropy can give us
more information on the phase structure of the
 holographic superconducting system at low temperatures and possibly of the
formation of FFLO states.

\begin{acknowledgments}
This work is supported partially by the NNSF of China and the
Shanghai Science and Technology Commission under Grant No.\
11DZ2260700. E.P is supported partially by ARISTEIA II action of
the operational programme education and long life learning which
is co-funded by the European Union (European Social Fund) and
National Resources.
\end{acknowledgments}

\end{document}